\documentclass[preprint,aps]{revtex4}

\newcommand{\be}{\begin{equation}}
\newcommand{\ee}{\end{equation}}
\newcommand{\bea}{\begin{eqnarray}}
\newcommand{\eea}{\end{eqnarray}}
\newcommand{\bd}{\begin{displaymath}}
\newcommand{\ed}{\end{displaymath}}

\newcommand{\ad }{{\hat{a}}^{\dagger}}

\newcommand{\ah }{ \hat{a}}

\begin{document}


\title{
Even and odd generalized hypergeometric coherent states }

\author{ Won Sang Chung }
\email{mimip4444@hanmail.net}

\author{ Mahouton Norbert Hounkonnou${}^b$}
\email{norbert.hounkonnou@cipma.uac.bj}

\author{ Sama Arjika${}^b$ }
\email{rjksama2008@gmail.com}

\affiliation{
Department of Physics and Research Institute of Natural Science, College of Natural Science, Gyeongsang National University, Jinju 660-701, Korea
}
\vspace{1cm}

\affiliation{
${}^b$International Chair of Mathematical Physics
and Applications
(ICMPA-UNESCO Chair), University of
Abomey-Calavi,
072 B. P.: 50 Cotonou, Republic of Benin}

\date{\today}

\begin{abstract}
In this paper, we   investigate a large class of    generalized hypergeometric states
$|p,q,z\rangle$, depending on a complex variable $z$ and two sets of parameters,
$(a_1,\cdots,a_p)$ and $(b_1,\cdots,b_q)$.  
    Even and  odd generalized hypergeometric 
states $|p,q,z\rangle_e$ and $|p,q,z\rangle_o$ are also defined and analyzed. 
The  moment problem is solved by 
the Mellin transform techniques.  For particular values 
of $p$ and $q$,  the photon-counting statistics, quantum 
optical properties and geometry of  these states are discussed.

\end{abstract}

\maketitle

\section{Introduction}
Coherent states are at the heart of important investigations in  physics and mathematical physics since their potential applications were highlighted in  \cite{Klauder&Skagerstam}, specifically in  quantum optics 
\cite{A.Perelomov}. They were introduced by Schr\"odinger
 \cite{E.Schrodinger}, but also studied by   Glauber \cite{R.Glauber} as the eigenstates 
of the boson annihilation operator $\ah$, i.e., 
\be
\ah |z\rangle=z|z\rangle
\ee
where $[\ah,\ad]=I,\,z$ is a complex number with conjugate $\bar{z};$  $|z\rangle$ are the normalized states given by
\be
|z\rangle  = e^{-\frac{|z|^2}{2}} \sum_{n=0}^{\infty} \frac{z^n }{ \sqrt{ n!}} |n\rangle,
\ee
where $|n\rangle$ is an element of the Fock space $\equiv \{|n\rangle,\,n=0,1,\cdots\}.$
 The  coherent states based on the
Heisenberg-Weyl group
  were extended for a number of Lie groups with
square integrable representations.
 For more details on  their  applications in physics, see also
\cite{W.Zhang&D.Feng&R.Gilmore,S.Ali&J.Antoine&J.Gazeau,Klauder,Littlejohn}.

Later on, Klauder et {\it al} \cite{Klauder&PensonS} exposed
a  general method for constructing holomorphic
coherent states of the form
\be
\label{label}
|z\rangle  = \frac{1}{\sqrt{N(|z|^2)}} \sum_{n=0}^{\infty} \frac{z^n }{ \sqrt{ \rho(n)}} |n\rangle
\ee
where $z $ is a complex variable, $\rho(n)$ is a set of strictly positive
parameters and the states $|n\rangle $ form an orthonormal
basis. The normalization function
is given by 
\be
N(x) = \sum_{n=0}^{\infty} \frac{x^n }{ \rho(n)},\quad  x = |z|^2.
\ee
Its radius of convergence determines the
domain of definition of the states in (\ref{label}). 

The idea of the even and odd coherent states was first introduced 
by Dodonov, Malkin and Man'ko \cite{V.Dodonov&I.Malkin&V.Man'ko}. Nieto and Traux 
\cite{M.Nieto&D.Traux} showed that these  states  are a special set 
of nonclassical states. Their properties 
 were studied by some authors \cite{V.Buzek&A.Viiella-Barranco&P.Knight,M.Hillery}. The even coherent states 
look like squeezed vacuum states  \cite{J.Hollenhorst}, because 
they are the superposition of the photon number states 
with an even number of quanta. Thus, the even coherent
 light can be used in interferometric gravitational wave 
detectors to give the same effort of increasing the sensitivity of 
these devices. The photon statistics of one-mode even and 
odd coherent states possesses the nonclassical property of light.

The even and odd coherent states were generalized to the case of even and odd nonlinear
coherent states \cite{Mancini,Sivakumar}
defined as the eigenstates of the 
operator $ f(N) \ah^2 $. In this case, the superposition of even number of
states yields the even nonlinear coherent states,  while the superposition of odd number of
states determines the odd nonlinear coherent states. For $f(N) =1 $, the even nonlinear 
coherent state becomes the even coherent
state and the odd nonlinear coherent state gives the odd coherent state. Depending
on the expression of $ f(N)$, the even and odd nonlinear coherent states may exhibit various
nonclassical features. Thus, the squeezed vacuum state and the squeezed
first excited state of the harmonic oscillator can be interpreted as even and odd nonlinear
coherent states, with $ f(N) = 1 / ( 1 + N) $ and $ f(N) = 1/ (2 + N), $ respectively. 

In this paper, we   investigate a large class of    generalized hypergeometric states
$|p,q,z\rangle$, depending on a complex variable $z$ and two sets of parameters,
$(a_1,\cdots,a_p)$ and $(b_1,\cdots,b_q)$.  Besides, we define and analyze
    even and  odd generalized hypergeometric 
states $|p,q,z\rangle_e$ and $|p,q,z\rangle_o.$ 
We solve the  moment problem using
the Mellin transform techniques.  For particular values 
of $p$ and $q$,  we discuss the photon-counting statistics, quantum 
optical properties and geometry of  these states.

The paper is organized as follows. In Section II, we recall   known    generalized 
hypergeometric states, define and analyze the even and  odd generalized hypergeometric states.
  In Section III,  we discuss their overcompleteness properties.  The associated 
moment problem is  solved.  Photon-counting statistics is discussed 
in the Section IV. We end with the study of  quantum optical and thermodynamical properties and geometry of these states
for particular relevant values of $ p, q$ in  Section V.

\section{Even and odd generalized hypergeometric states}
Let us consider the  Fock space
$\mathcal{F}:=\big\{|n\rangle,\;\,\;\,n=0,1,2,\cdots\big\}$
with  the orthogonality and completeness conditions
\be 
\langle n|m\rangle=\delta_{n,m}\quad\mbox{ and }\quad \sum_{n=0}^\infty|n\rangle\langle n|=I
\ee
such that the state $|n\rangle $ is obtained by acting the creation operator  $\ad $ of boson algebra on the vacuum $|0\rangle $ repeatedly as follows:
\be
|n\rangle = \frac{(\ad)^n }{\sqrt{ n!}}|0 \rangle.
\ee
The   actions  of the operators $\ah$ and $\ad$ on the state  $|n\rangle$  are given by
\be 
\label{actions:gen}
 \ah|n\rangle =\sqrt{n}\,|n-1\rangle \quad \mbox{and} \quad 
\ad|n\rangle =\sqrt{n+1}\,|n+1\rangle.
\ee
As introduced in \cite{Appl},   the generalized hypergeometric states (GHS) are defined as
\be
\label{apap}
|p,q,z\rangle = | a_1 , \cdots, a_p ; b_1, \cdots, b_q ; z\rangle= N(|z|^2;p,q) \sum_{n=0}^{\infty} \frac{z^n }{ \sqrt{ {}_p\rho_q(n)}} |n\rangle,
\ee
where the normalization function $N(x;p,q) $ is  afforded by the generalized
hypergeometric function:
\be
N(x;p,q) =[ {}_p F_q (a_1, \cdots , a_p; b_1, \cdots , b_q; x)]^{-1/2},
\ee
with
\be
 {}_p F_q (a_1, \cdots , a_p; b_1, \cdots , b_q; x) = \sum_{n=0}^{\infty}  \frac{(a_1)_n \cdots (a_p )_n } {(b_1)_n \cdots (b_q)_n }\frac{x^n}{n!}
 \ee 
and the strictly positive parameter function
\be
{}_p\rho_q(n)= n! \frac{ (b_1)_n \cdots (b_q)_n }{(a_1)_n \cdots (a_p )_n },
\ee
where the    Pochhammer-symbol $(a)_n$ is    defined as \cite{ASK} 
\bd
(a)_n = a( a+1) \cdots ( a + n-1) , ~~~n \ge 1, ~~~~ (a)_0 =1.
\ed
The function ${}_p F_q (a_1, \cdots , a_p; b_1, \cdots , b_q; x)$ converges in the following cases: for 
\bea
\label{sa:ma}
\mbox{ any } z \mbox{ if } \quad p < q+1,\\
\label{sa:ama}
|z| < 1 \quad \mbox{if}\quad p=q+1,\\
|z|=1 \quad \mbox{if}\quad p=q+1,\quad \eta=1,\\
|z|=1, \quad z\neq 1 \quad \mbox{if}\quad p=q+1,\quad 0\leq \eta\leq 1,
\eea
where 
\be
\eta=Re\left(\sum_{j=1}^pa_j-\sum_{j=1}^qb_j\right).
\ee
In all other cases, it diverges \cite{Marichev}.
The GHS depend on the complex
variable $z$ and on the two sets of sequences $ (a_1, \cdots , a_p)$
and $ (b_1, \cdots,  b_q)$. We suppose that the parameters  
$a_1, \cdots , a_p,b_1, \cdots , b_q$ are  non zero, non 
negative real parameters. The parameter functions ${}_p\rho_q(n)$ is a real and strictly positive and 
\be
\label{positive}
\frac{(b_1+n)(b_2+n)\cdots(b_q+n)}{(a_1+n)(a_2+n)\cdots(a_p+n)}>0,\quad n=0,\,1,\,2,\,\cdots.
\ee

Let us define  the even generalized hypergeometric states (EGHS) and the odd generalized hypergeometric states (OGHS)  as follows:
\bd
|p,q,z\rangle_e : = \frac{1}{\sqrt{2}}N'_e ( x; p,q) ( |p,q,z\rangle + |p,q,-z\rangle )
\ed
and
\be
|p,q,z\rangle_o  := \frac{1}{\sqrt{2}} N'_o ( x; p,q) ( |p,q,z\rangle - |p,q,-z\rangle ),
\ee
respectively, where the normalization functions $N'_e ( x; p,q)$ and $N'_o ( x; p,q)$ are given by
\bd
N'_e ( x; p,q) = \left[ \frac{ {}_p F_q (a_1, \cdots , a_p; b_1, \cdots , b_q; x)}{{}_p C_q (a_1, \cdots , a_p; b_1, \cdots , b_q; x)} \right]^{1/2}
\ed
and
\be
N'_o ( x; p,q) = \left[ \frac{ {}_p F_q (a_1, \cdots , a_p; b_1, \cdots , b_q; x)}{{}_p S_q (a_1, \cdots , a_p; b_1, \cdots , b_q; x)} \right]^{1/2}.
\ee
 The even generalized hypergeometric function is defined as
\be
 {}_p C_q (a_1, \cdots , a_p; b_1, \cdots , b_q; x) = \sum_{n=0}^{\infty}  \frac{(a_1)_{2n} \cdots (a_p )_{2n} } { (b_1)_{2n} \cdots (b_q)_{2n} } \frac{x^{2n}}{(2n)!}
 \ee 
while   the odd generalized hypergeometric function is provided by 
 \be
 {}_p S_q (a_1, \cdots , a_p; b_1, \cdots , b_q; x) = \sum_{n=0}^{\infty}  \frac{(a_1)_{2n+1} \cdots (a_p )_{2n+1} } { (b_1)_{2n+1} \cdots (b_q)_{2n+1} } \frac{x^{2n+1}}{(2n+1)!}.
 \ee 
Therefore, the even and odd generalized hypergeometric function satisfy 
\be
{}_p F_q (a_1, \cdots , a_p; b_1, \cdots , b_q; x) = {}_p C_q (a_1, \cdots , a_p; b_1, \cdots , b_q; x)  + {}_p S_q (a_1, \cdots , a_p; b_1, \cdots , b_q; x)  
 \ee
 and 
 \bd
 [{}_p C_q (a_1, \cdots , a_p; b_1, \cdots , b_q; x)  ]^2 - [ {}_p S_q (a_1, \cdots , a_p; b_1, \cdots , b_q; x) ]^2 
 \ed
 \be
 = {}_p F_q (a_1, \cdots , a_p; b_1, \cdots , b_q; x) {}_p F_q (a_1, \cdots , a_p; b_1, \cdots , b_q; -x).
 \ee
 The EGHS and OGHS can be rewritten as 
 \be
\label{sama:sama:sama}
 |p,q, z\rangle_e = \frac{1}{ \sqrt{ {}_p C_q (a_1, \cdots , a_p; b_1, \cdots , b_q;|z|^2)  } } 
\sum_{n=0}^{\infty} \sqrt{ \frac{(a_1)_{2n} \cdots
 (a_p )_{2n} } {(2n)! (b_1)_{2n} \cdots (b_q)_{2n} } }\,z^{2n} |2n\rangle
 \ee
and
 \be
\label{sama:sama:sa}
 |p,q, z\rangle_o = \frac{1}{ \sqrt{ {}_p S_q (a_1, \cdots , a_p; b_1, \cdots , b_q; |z|^2)  } } 
\sum_{n=0}^{\infty} \sqrt{ \frac{(a_1)_{2n+1} \cdots
 (a_p )_{2n+1} } {(2n+1)! (b_1)_{2n+1} \cdots (b_q)_{2n+1} } }\,z^{2n+1} |2n+1\rangle.
 \ee
 From  (\ref{sama:sama:sama}) and (\ref{sama:sama:sa}),   the EGHS and OGHS satisfy the orthonormality condition
\be
{}_e \langle z,p,q |p,q, z\rangle_o = 0
\ee
and the action of the annihilation operator on them yields, respectively,
\bd
\ah |p, q, z\rangle_e = z \sqrt{ \frac{ a_1 \cdots a_p }{ b_1 \cdots b_q }}
 \sqrt{ \frac{ {}_p S_q (a_1 +1 , \cdots , a_p +1 ; b_1 +1 , \cdots , b_q +1 ; |z|^2)}{ 
{}_p C_q (a_1 , \cdots , a_p ; b_1 , \cdots , b_q ;|z|^2)}}\, |p, q, z\rangle_o
\ed
and
\be
\ah |p, q, z\rangle_o = z \sqrt{ \frac{ a_1 \cdots a_p }{ b_1 \cdots b_q }} 
\sqrt{ \frac{ {}_p C_q (a_1 +1 , \cdots , a_p +1 ; b_1 +1 , \cdots , b_q +1 ; |z|^2)}{ 
{}_p S_q (a_1 , \cdots , a_p ; b_1 , \cdots , b_q ; |z|^2)}}\, |p, q, z\rangle_e,
\ee
where the identities
\bd
(a)_n = a ( a +1)_{n-1}, ~~~ (a)_n = a ( a +1) (a+2)_{n-2}
\ed
are used.
The EGHS and OGHS are also continuous in their label $z$. 
\section{ Completeness}
The overcompleteness property
consists of finding a positive function ${}_pW_q(|z|^2)$ such that
\be
\label{pro:mo}
\frac{1}{\pi}\int\int_{D}d^2z|p,q,z\rangle {}_pW_q(|z|^2)\langle z,p,q|=I=\sum_{n=0}^\infty |n\rangle\langle n|,
\ee
where $D$ is a disc in the complex plane centered at the origin, of radius $R$ and $d^2z=|z|d|z|d\theta.$
By substituting $z=\sqrt{x}\,e^{i\theta}$ in this equation, and evaluating the integral over $\theta$ in the l.h.s of (\ref{pro:mo}),  we finally arrive at
\be
\label{moment:pro}
 \int_{0}^R x^n\left[\frac{{}_pW_q(x)}{N^2(x;p,q)}\right]dx={}_p\rho_q(n),\quad  x=|z|^2.
\ee
Hence, (\ref{moment:pro}) is  
then the power moments of the unknown 
function $ {}_pW_q(x)$ and 
is the Stieltjes moment problem for $R=\infty$ or the Haussdorf moment problem  
  for  $  R<\infty.$ These are classical 
mathematical problems on which an extensive and mathematically oriented literature
exists, see for instance  \cite{Akhiezer,Tamarkin,Simon} and references therein.
Under the considerations (\ref{sa:ma}) and (\ref{sa:ama}),  the moment problem   (\ref{moment:pro}) 
takes the form
\be
\label{sama:even}
 \int_{0}^R x^n{}_p\omega_q(x)dx={}_p\rho_q(n),
\ee
where  ${}_p\omega_q(x)={}_pW_q(x)/N^2(x;p,q).$
Its solution   can be obtained  by
using Mellin transform techniques \cite{KlauderJR,QuesneC}.  By replacing $n$
 by  $s-1$ in  (\ref{sama:even}),  the distribution  ${}_p\omega_q(x)$ and the parameter function ${}_p\rho_q(s-1)$
become a Mellin transform related pair \cite{grad,Marichev}:
\be
 \int_{0}^\infty x^{s-1} {}_p\omega_q(x) dx=\Gamma(s)\frac{\prod_{i=1}^q\Gamma(b_i-1+s)}{\prod_{i=1}^p\Gamma(a_i-1+s)},\quad  x=|z|^2.
\ee
For $p< q+1$ and 
$p=q+1,\;\eta >1$, the distribution   ${}_p\omega_q(x)$ is given by \cite{Appl}
\bea
{}_p\omega_q(x)&=&{}_p\omega_q(x;a_1, \cdots,a_p;b_1,\cdots,b_q)\cr
&=&\frac{\Gamma(a_1)\cdots\Gamma(a_p)}{\Gamma(b_1)\cdots\Gamma(b_p)} \,G_{p,q+1}^{q+1,0}\left(x\,\Bigg|\begin{array}{c}a_1-1, \cdots,a_p-1\\
b_1-1,\cdots,b_q-1,0\end{array}\right)
\eea
and the weight function  ${}_pW_q(x)$ is furnished by the expression
\bea
{}_pW_q(x)&=&{}_pW_q(x;a_1, \cdots,a_p;b_1,\cdots,b_q)\cr
&=&\frac{\Gamma(a_1)\cdots\Gamma(a_p)}{\Gamma(b_1)\cdots\Gamma(b_p)}{}_pF_q(a_1, \cdots,a_p;b_1,\cdots,b_q;x)\cr
&&\times \; G_{p,q+1}^{q+1,0}\left(x\,\Bigg|\begin{array}{c}a_1-1, \cdots,a_p-1\\
b_1-1,\cdots,b_q-1,0\end{array}\right)
\eea
where $G$  is the Meijer function \cite{Marichev,grad}.\\
For the EGHS and the OGHS, the moment problem (\ref{sama:even}) takes the form
\be
\label{mo:even}
 \int_{0}^\infty x^{2n}{}_p\omega_q^e(x)dx={}_p\rho_q(2n),
\ee
and 
\be
\label{mo:odd}
 \int_{0}^\infty x^{2n+1}{}_p\omega_q^o(x)dx={}_p\rho_q(2n+1),
\ee
respectively, where ${}_p\omega_q^e(x)={}_pW_q^e(x)/{}_pC_q(x;p,q)$ and ${}_p\omega_q^o(x)={}_pW_q^o(x)/{}_pS_q(x;p,q).$\\
%
In particular, 
\begin{enumerate}
\item For $p=0=q,$   the states $|0,0,z\rangle_e=|z\rangle_e$ and $|0,0,z\rangle_o=|z\rangle_o$ related to
${}_0\rho_0(n)=n!$   have the normalization functions  given by \cite{Sivakumar}
\be
  {}_0C_0(-;-;x)=\cosh x,\quad {}_0S_0(-;-;x)=\sinh x
\ee
and the  corresponding weight functions  are
\be
{}_0\omega_0^e(x)={}_0\omega_0^e(x;-;-)\equiv{}_0\omega_0^o(x)={}_0\omega_0^o(x;-;-)=e^{-x}.
\ee
\item For $p=0,\;q=1,$ the states $|0,1,z \rangle_e =|-,b,z \rangle_e $  and 
$|0,1,z \rangle_o=|-,b,z \rangle_o $  related to  ${}_0\rho_1(n)=n!(b)_n$ for $b>0$ due to  (\ref{positive})
lead to the normalization functions
 \be
 {}_0C_1(-;b;x)={}_0F_3\left( \begin{array}{c}-\\\frac{1}{2},\frac{b}{2},\frac{b+1}{2}
\end{array}\Bigg|\;\frac{x^2}{16}\right),\quad {}_0S_1(-;b;x)=\frac{4x}{b}{}_0F_3\left( \begin{array}{c}-\\\frac{3}{2},\frac{b+1}{2},\frac{b+2}{2}
\end{array}\Bigg|\;\frac{x^2}{16}\right)
\ee
and the weight functions by (\cite{OberhettingerF}, p 196, formula (5.39)) are reduced to
\be
{}_0\omega_1^e(x)={}_0\omega_1^e(x;-,b)\equiv {}_0\omega_1^o(x)={}_0\omega_1^o(x;-,b)=\frac{2 }{\Gamma(b)}\,x^{\frac{b-1}{2}}K_{b-1}(2\sqrt{x}),
\ee
where $K_{\alpha}(\cdot)$ is the modified Bessel function \cite{OberhettingerF}.\\
\item For  $p=1,\,q=0,$
  the states $|1,0,z \rangle_e =|a,-,z \rangle_e $  and 
$|1,0,z \rangle_o=|a,-,z \rangle_o $  originate
 from ${}_1\rho_0(n)=n!/(a)_n,$ with $a >0$  due
to (\ref{positive}). The normalization functions are  given by
\be
{}_1C_0(a;-;x)={}_2F_1\left( \begin{array}{c}\frac{a}{2},\frac{a+1}{2}\\
\frac{1}{2}\end{array}\Bigg|\;x^2\right),\quad
{}_1S_0(a;-;x)=ax\,{}_2F_1\left( \begin{array}{c}\frac{a+1}{2},\frac{a+2}{2}\\
\frac{3}{2}\end{array}\Bigg|\;x^2\right).
\ee
The   corresponding weight functions by (\cite{OberhettingerF}, p 195, formula (5.35)) are expressed as follows:
\be 
{}_1\omega_0^e(x)={}_1\omega_0^e(x;a ;-)\equiv{}_1\omega_0^o(x)={}_1\omega_0^o(x;a ;-)=(a-1)(1-x)^{a-2}.
\ee 
\item For  $p=1= q,$
  the states $|1,1,z \rangle_e =|a,b,z \rangle_e $  and 
$|1,1,z \rangle_o=|a,b,z \rangle_o $  originate
 from ${}_1\rho_1(n)=n!(b)_n/(a)_n,$ with $a,\,b>0$  due
to (\ref{positive}). The normalization functions are  given by
\be
{}_1C_1(a;b;x)={}_2F_3\left( \begin{array}{c}\frac{a}{2},\frac{a+1}{2}\\
\frac{1}{2},\frac{b}{2},\frac{b+1}{2}\end{array}\Bigg|\;\frac{x^2}{4}\right),\quad
{}_1S_1(a;b;x)=\frac{2ax}{b}\,{}_2F_3\left( \begin{array}{c}\frac{a+1}{2},\frac{a+2}{2}\\
\frac{3}{2},\frac{b+1}{2},\frac{b+2}{2}\end{array}\Bigg|\;\frac{x^2}{4}\right).
\ee
The   corresponding weight functions by (\cite{OberhettingerF}, p 197, formula (5.46)) can be expressed in terms of the Whittaker function as follows:
\be
{}_1\omega_1^e(x)={}_1\omega_1^e(x;a ;b)\equiv{}_1\omega_1^o(x)={}_1\omega_1^o(x;a ;b)
=\frac{\Gamma(a) }{\Gamma( b) }\,
e^{ -\frac{x}{2}}\,W_{\frac{1+b}{2}-a,-\frac{b}{2}}(x),
\ee
where $W_{\alpha,\beta}(\cdot)$ is the Whittaker function \cite{OberhettingerF}.
\item For  $p=2,\; q=1,$
  the states $|2,1,z \rangle_e =|a_1,a_2,b,z \rangle_e $  and 
$|2,1,z \rangle_o=|a_1,a_2,b,z \rangle_o $  originate
 from ${}_2\rho_1(n)=n!(b)_n/(a_1,a_2)_n,$ with $a_1,\,a_2,\,b>0$  due
to (\ref{positive}). The normalization functions are  given in terms of hypergeometric function ${}_4F_3:$
\bea
{}_2C_1(a_1,a_2;b;x)={}_4F_3\left( \begin{array}{c}\frac{a_1}{2},\frac{a_2}{2},\frac{a_1+1}{2},\frac{a_2+1}{2}\\
\frac{1}{2},\frac{b}{2},\frac{b+1}{2}\end{array}\Bigg|\; x^2 \right),\\
{}_2S_1(a_1,a_2;b;x)=\frac{a_1a_2x}{b}\,{}_4F_3\left( \begin{array}{c}\frac{a_1+1}{2},\frac{a_2+1}{2},\frac{a_1+2}{2},\frac{a_2+2}{2}\\
\frac{3}{2},\frac{b+1}{2},\frac{b+2}{2}\end{array}\Bigg|\; x^2 \right)
\eea
while the   corresponding weight functions by (\cite{OberhettingerF}, p 198, formula (5.50)) are obtained in terms of  ${}_2F_1:$
\bea
{}_2\omega_1^e(x)&=&{}_2\omega_1^e(x;a_1,a_2 ;b)\equiv{}_2\omega_1^o(x)={}_2\omega_1^o(x;a_1,a_2 ;b)\cr
&=&\frac{\Gamma(a_1)\Gamma(a_2) }{\Gamma( b) \Gamma(a_1+a_2-b-1)}\,
(1-x)^{a_1+a_2-b-2}\,{}_2F_1\left( \begin{array}{c} a_1-b, a_2-b\\
a_1+a_2-b-1\end{array}\Bigg|\; 1-x  \right).
\eea
If $b=a_1$ or $b=a_2,$ we recover the results of the previous example for $p=1,\,q=0$.
\end{enumerate}
For any positive parameters   $p,\,q$, the weight 
functions solving the moment problems (\ref{mo:even}) and (\ref{mo:odd}) are provided by the expression
\bea
{}_p\omega_q^e(x)&=&{}_p\omega_q^e(x;a_1, \cdots,a_p;b_1,\cdots,b_q)\equiv{}_p\omega_q^o(x)={}_p\omega_q^o(x;a_1, \cdots,a_p;b_1,\cdots,b_q)\cr\cr
&=&\frac{\Gamma(a_1)\cdots\Gamma(a_p)}{\Gamma(b_1)\cdots\Gamma(b_p)}
  G_{p,q+1}^{q+1,0}\left(x\;\Bigg|\begin{array}{c}a_1-1, \cdots,a_p-1\\
b_1-1,\cdots,b_q-1,0\end{array}\right).
\eea
Provided these results, we readily obtain the  completeness relations corresponding to the
families of the  states $|p,q,z\rangle_e$ and $|p,q,z\rangle_o.$

\section{ Photon-counting statistics}
In this section, we compute the expectation value of $(\ad)^s \ah^r$ in the generalized coherent states 
$|p,q,z \rangle_e $  and $|p,q,z \rangle_o $ and deduce the corresponding mandel parameter.

From the actions  of the operators $\ah$ and 
$\ad$ on the state  $|n\rangle$  in  (\ref{actions:gen}) we deduce that
\be 
\label{ah:ad}
 \ah^r|n\rangle =\sqrt{\frac{n!}{(n-r)!}}\,|n-r\rangle, \quad 0\leq r\leq n
\ee
and
\be 
\label{ah:ad1}
 (\ad)^s|n\rangle=\sqrt{\frac{(n+s)!}{n!}}\,|n+s\rangle.
\ee
The GHCS defined in (\ref{apap}) have the Fock representation
\be
\langle n|p,q,z\rangle=\frac{z^n}{\sqrt{{}_p\rho_q(n)\,{}_p F_q (a_1, \cdots , a_p; b_1, \cdots , b_q; |z|^2)}}
\ee
from which the photon number distribution follows as:
\be
\mathcal{P}_{|p,q,z\rangle}(n)=|\langle n|p,q,z\rangle|^2=\frac{x^n}{{}_p\rho_q(n)\,{}_p F_q (a_1, \cdots , a_p; b_1, \cdots , b_q; x)},\quad x=|z|^2.
\ee
Besides, we can prove the following statement.

 {\bf Proposition 1.} \it 
 The expectation value of $(\ad)^s \ah^r$ in the generalized coherent states 
$|p,q,z \rangle_e $  and $|p,q,z \rangle_o $ are given by
\be 
\label{sama:abel:ma}
  \langle(\ad)^s \,\ah^r\rangle_e
=\bar z^sz^r{}_p{\cal S}_q^{(s,r)}(|z|^2),\quad s,\,r=0,\;1,\;2,\cdots
\ee
and
\be 
\label{sama:abelma}
  \langle(\ad)^s \,\ah^r\rangle_o
=\bar z^sz^r{}_p\tilde{{\cal S}}_q^{(s,r)}(|z|^2),\quad s,\,r=0,\;1,\;2,\cdots,
\ee
respectively,
where
\be 
\label{abel:saa}
{}_p{\cal S}_q^{(s,r)}(x)=\frac{1}{{}_pC_q(a_1, \cdots,a_p;b_1,\cdots,b_q;x)}\sum_{m=0}^\infty 
 \sqrt{\frac{(2m+r)!\,(2m+s)!}{{}_p\rho_q(2m+s)\,{}_p\rho_q(2m+r)}}\,\frac{x^{2m}}{(2m)!},\quad x=|z|^2.
\ee
and
\be 
\label{abelaa}
{}_p\tilde{{\cal S}}_q^{(s,r)}(x)=\frac{1}{{}_pS_q(a_1, \cdots,a_p;b_1,\cdots,b_q;x)}\sum_{m=0}^\infty  \sqrt{\frac{(2m+r+1)!(2m+s+1)!}{{}_p\rho_q(2m+s+1){}_p\rho_q(2m+r+1)}}\,\frac{x^{2m+1}}{(2m+1)!}.
\ee
In particular, for the EGHS, we have
\be 
  \langle(\ad)^r \,\ah^r\rangle_e= \frac{x^{r}}{{}_pC_q(a_1, \cdots,a_p;b_1,\cdots,b_q;x)}\left(\frac{d}{dx}\right)^r{}_pC_q(a_1, \cdots,a_p;b_1,\cdots,b_q;x)\label{especatation}
\ee 
which is equivalent to
\be
  \langle(\ad)^r \,\ah^r\rangle_e=
=x^{r}\frac{(a_1,\cdots,a_p)_{r}}{(b_1,\cdots,b_q)_{r}} \frac{{}_pC_q(a_1+r, \cdots,a_p+r;b_1+r,\cdots,b_q+r;x)}{{}_pC_q(a_1, \cdots,a_p;b_1,\cdots,b_q;x)}
\ee
if  $r$ is even, or
\be 
  \langle(\ad)^r \,\ah^r\rangle_e
=x^{r}\frac{(a_1,\cdots,a_p)_{r}}{(b_1,\cdots,b_q)_{r}} \frac{{}_pS_q(a_1+r, \cdots,a_p+r;b_1+r,\cdots,b_q+r;x)}{{}_pC_q(a_1, \cdots,a_p;b_1,\cdots,b_q;x)}\label{espeatation}
\ee 
if $r$ is odd.\\
For the OGHS, we have
\be 
  \langle(\ad)^r \,\ah^r\rangle_o= \frac{x^{r}}{{}_pS_q(a_1, \cdots,a_p;b_1,\cdots,b_q;x)}\left(\frac{d}{dx}\right)^r{}_pS_q(a_1, \cdots,a_p;b_1,\cdots,b_q;x)
\ee 
which is equibalent to
\be 
  \langle(\ad)^r \,\ah^r\rangle_o
=x^{r}\frac{(a_1,\cdots,a_p)_{r}}{(b_1,\cdots,b_q)_{r}} \frac{{}_pS_q(a_1+r, \cdots,a_p+r;b_1+r,\cdots,b_q+r;x)}{{}_pS_q(a_1, \cdots,a_p;b_1,\cdots,b_q;x)}
\ee 
if $r$ is  even, or
\be 
  \langle(\ad)^r \,\ah^r\rangle_o
=x^{r}\frac{(a_1,\cdots,a_p)_{r}}{(b_1,\cdots,b_q)_{r}} \frac{{}_pC_q(a_1+r, \cdots,a_p+r;b_1+r,\cdots,b_q+r;x)}{{}_pS_q(a_1, \cdots,a_p;b_1,\cdots,b_q;x)}\label{especatation1}
\ee 
  if $r$ is  odd,
where $(a_1,\cdots,a_p)_n:=(a_1)_n(a_2)_n\cdots(a_p)_n.$
\rm
\\
{\bf Proof.}  Indeed, for $s,r=0,\;1,\;2,\cdots$, we have 
\bea
 \langle(\ad)^s \,\ah^r\rangle_e:&=&{}_e\langle z,p,q |(\ad)^s \, \ah^r|p,q,z\rangle_e
\cr
&=&\frac{1}{  {}_pC_q(a_1, \cdots,a_p;b_1,\cdots,b_q;|z|^2)}\cr\cr
&\times&\sum_{m=0}^\infty\sum_{2n=r}^\infty  \sqrt{\frac{(2n)!\,(2n-r+s)!}{{}_p\rho_q(2m)\,{}_p\rho_q(2n)(2n-r)!\,(2n-r)!}}\,\bar z^{2m}z^{2n} \langle 2m|2n+s-r\rangle 
\cr\cr
&=&\frac{1}{  {}_pC_q(a_1, \cdots,a_p;b_1,\cdots,b_q;|z|^2)}\cr\cr
&\times&\sum_{2n=r}^\infty 
 \sqrt{\frac{(2n)!\,(2n-r+s)!}{{}_p\rho_q(2n+s-r)\,{}_p\rho_q(2n)(2n-r)!\,(2n-r)!}}\,\bar z^{2n+s-r}z^{2n}
\cr\cr
&=&\frac{\bar z^sz^r}{ {}_pC_q(a_1, \cdots,a_p;b_1,\cdots,b_q;|z|^2)}\sum_{n=0}^\infty  \sqrt{\frac{(2n+r)!(2n+s)!}{{}_p\rho_q(2n+s)\,{}_p\rho_q(2n+r)}}\frac{|z|^{4n}}{(2n)!}.
\eea
In the special case, when $s=r$, we have 
\bea
 \langle(\ad )^r\,\ah^r\rangle_e&=&\frac{x^{r}}{ {}_pC_q(a_1, \cdots,a_p;b_1,\cdots,b_q;x^2)}\sum_{n=0}^\infty   \frac{(2n+r)!}{{}_p\rho_q(2n+r)}\frac{x^{2n}}{(2n)!}
\cr&=& \frac{x^r}{ {}_pC_q(a_1, \cdots,a_p;b_1,\cdots,b_q;x)}\sum_{r=2n}^\infty   \frac{(2n)!}{{}_p\rho_q(2n)}\frac{x^{2n-r}}{(2n-r)!}
\cr&=& \frac{x^r}{ {}_pC_q(a_1, \cdots,a_p;b_1,\cdots,b_q;x)}\left(\frac{d}{dx}\right)^r{}_pC_q(a_1, \cdots,a_p;b_1,\cdots,b_q;x),
\eea
where $ x=|z|^2.$
In the same way, we compute $\langle(\ad )^s\,\ah^r\rangle_o,$ which achives the proof.\\
In particular, 
\begin{enumerate}
\item for $r=1,$ we deduce   the expectation value of the number operator as
\be 
 \langle N\rangle_e\equiv\langle \ad  \,\ah\rangle_e=
x\left[ \frac{\prod_{i=1}^p a_i }{\prod_{i=1}^q b_i } \right]   
\frac{{}_pS_q (a_1+1 , \cdots,a_p+1 ;b_1+1 ,\cdots,b_q+1 ;x)}{{}_pC_q(a_1, \cdots,a_p;b_1,\cdots,b_q;x)} 
\ee
and
\be 
 \langle N\rangle_o\equiv\langle \ad  \,\ah\rangle_o=x\left[ \frac{\prod_{i=1}^p a_i }{
\prod_{i=1}^q b_i } \right]   \frac{{}_pC_q (a_1+1 , \cdots,a_p+1 ;b_1+1 ,\cdots,b_q+1 ;x)}{{}_pS_q(a_1, \cdots,a_p;b_1,\cdots,b_q;x)},\, \,x=|z|^2;
\ee
\item for $r=2$, the computation of
the expectation value of the square of the number operator $N^2=(\ad)^2\ah^2+\ad\ah$  turns to be
\bea
\langle N^2 \rangle_e &=& x^2 \left[ \frac{\prod_{i=1}^p a_i (a_i +1)}{
\prod_{i=1}^q b_i (b_i +1 ) } \right] \frac{  {}_p C_q (a_1 +2 , \cdots , a_p +2 ; b_1 +2 , \cdots , b_q +2 ; x)}{ {}_p C_q (a_1 , \cdots , a_p ; b_1 , \cdots , b_q ; x)}\cr
&+& x \left[ \frac{\prod_{i=1}^p a_i }{\prod_{i=1}^q b_i } \right] 
\frac{  {}_p S_q (a_1 +1 , \cdots , a_p +1 ; b_1 +1 , \cdots , b_q +1 ; x)}{ {}_p C_q (a_1 , \cdots , a_p ; b_1 , \cdots , b_q ; x)}
\eea
and
\bea
\langle N^2 \rangle_o &=& x^2 \left[ \frac{\prod_{i=1}^p a_i (a_i +1)}{\prod_{i=1}^q b_i (b_i +1 ) } \right]
 \frac{  {}_p S_q (a_1 +2 , \cdots , a_p +2 ; b_1 +2 , \cdots , b_q +2 ; x)}{ {}_p S_q (a_1 , \cdots , a_p ; b_1 , \cdots , b_q ; x)}\cr
&+& x \left[ \frac{\prod_{i=1}^p a_i }{\prod_{i=1}^q b_i } \right] \frac{  {}_p C_q (a_1 +1 ,
 \cdots , a_p +1 ; b_1 +1 , \cdots , b_q +1 ; x)}{ {}_p S_q (a_1 , \cdots , a_p ; b_1 , \cdots , b_q ; x)},\;\, x=|z|^2.
\eea
\end{enumerate}
Commonly, the photon-counting statistics of the coherent states is 
investigated by evaluating the Mandel parameter defined as  \cite{DeyFring}
\be
\label{mandel:par}
Q_i  := \frac{ \langle N^2 \rangle_i  -  \langle N \rangle_i^2 }{ \langle N \rangle_i} -1,
\ee
where
\be
\langle N \rangle_i = {}_i\langle z | N | z \rangle_i
\ee
The coherent state for which $ Q_i  =0, \,Q_i < 0 $ and $ Q_i > 0 $
corresponds to Poissonian, sub-Poissonian (non-classical) and super-Poissonian state  \cite{mandel1,zhangetal,jpjpg}, respectively . \\
Since the EGHS and the OGHS are orthogonal, we get
\be
\langle \ah \rangle_e = \langle \ah \rangle_o =0= \langle \ad \rangle_e = \langle \ad \rangle_o
\ee
where
\bd
\langle \hat{X}  \rangle_{e(o)} = {}_{e(o)} \langle p,q, z| \hat{X} | p,q,z\rangle_{e(o)}.
\ed
The Mandel parameter (\ref{mandel:par}) is reduced to
\bea
Q_e&=&x  \left\{ \frac{\prod_{i=1}^p   (a_i +1)}{\prod_{i=1}^q  
 (b_i +1 ) }  \frac{  {}_p C_q (a_1 +2 , \cdots , a_p +2 ; b_1 +2 , \cdots , b_q +2 ; x)}{ {}_p S_q (a_1+1 , \cdots , a_p+1 ; b_1+1 , \cdots , b_q+1 ; x)}\right.\cr\cr
&-&  \left. \frac{\prod_{i=1}^p a_i }{\prod_{i=1}^q b_i }
  \frac{  {}_p S_q (a_1 +1 , \cdots , a_p +1 ; b_1 +1, \cdots , b_q +1 ; x)}{ {}_p C_q (a_1, \cdots , a_p ; b_1 , \cdots , b_q ; x)}\right\}
\eea
 and
\bea
Q_o&=&x  \left\{ \frac{\prod_{i=1}^p   (a_i +1)}{\prod_{i=1}^q   (b_i +1 ) }  
\frac{  {}_p S_q (a_1 +2 , \cdots , a_p +2 ; b_1 +2 , \cdots , b_q +2 ; x)}{ {}_pC_q (a_1+1, \cdots,
 a_p+1 ; b_1+1 , \cdots , b_q+1 ; x)}\right.\cr\cr
&-&  \left. \frac{\prod_{i=1}^p a_i }{\prod_{i=1}^q b_i }  \frac{  {}_p C_q (a_1 +1,
 \cdots , a_p +1 ; b_1 +1 , \cdots , b_q +1 ; x)}{ {}_p S_q (a_1, \cdots , a_p ; b_1, \cdots , b_q ; x)}\right\}.
\eea
The probability of finding $n$ quanta in the   EGHS and OGHS are given by
\be 
 {}_p\mathcal{P}_q^{even}(x,n):=| \langle 2n|p,q,z\rangle_e|^2= \frac{x^{2n}}{
{}_p\rho_q(2n)\, {}_pC_q(a_1, \cdots,a_p;b_1,\cdots,b_q;x)}, 
\ee
and
\be 
 {}_p\mathcal{P}_q^{odd}(x,n):=| \langle 2n+1|p,q,z\rangle_o|^2= \frac{x^{2n+1}}{
{}_p\rho_q(2n+1)\, {}_pS_q(a_1, \cdots,a_p;b_1,\cdots,b_q;x)},\quad x=|z|^2,
\ee
respectively. In particular when $p=0=q,$ we have
\be 
 {}_0\mathcal{P}_0^{even}(x,n):=| \langle 2n|-,-,z\rangle_e|^2= \frac{x^{2n}}{
 (2n)!\, \cosh x}, 
\ee
and
\be 
 {}_0\mathcal{P}_0^{odd}(x,n):=| \langle 2n+1|-,-,z\rangle_o|^2= \frac{x^{2n+1}}{
(2n+1)!\, \sinh x},\quad x=|z|^2,
\ee

 \section{ Quantum optical properties and geometry of the GHS $|p,q,z\rangle_i$ for $ p=1, q=0$}
 Now, we exploit the results issued from the   above section to derive
some quantum optical properties and describe the geometry of the states $|p,q,z\rangle_e $ and $|p,q,z\rangle_o $ for $ p=1, q=0.$ 

\subsection{Quantum optical properties}
 The expectation value of $(\ad)^s \ah^r$ in the coherent states 
$|1,0,z \rangle_e =|a,-,z \rangle_e $  and $|1,0,z \rangle_o=|a,-,z \rangle_o $ are given by
\be 
\label{sama:abel:ma}
  \langle(\ad)^s \,\ah^r\rangle_e
=\bar z^sz^r{}_1{\cal S}_0{(s,r)}(|z|^2),\quad s,r=0,\;1,\;2,\cdots
\ee
and
\be 
\label{sama:abelma}
  \langle(\ad)^s \,\ah^r\rangle_o
=\bar z^sz^r{}_1\tilde{{\cal S}}_0^{(s,r)}(|z|^2),\quad s,r=0,\;1,\;2,\cdots,
\ee
respectively.
In particular, 
\bea
  \langle(\ad)^r \,\ah^r\rangle_e=\frac{x^{r}}{{}_1C_0(a ;-;x)}\left(\frac{d}{dx}\right)^r{}_1C_0(a ;-; x)
=\left\{\begin{array}{ll}x^{r} (a)_{r} \frac{{}_1C_0(a +r ;-;x)}{{}_1C_0(a ;-;x)}&\quad\mbox{ if }\;r \;\mbox{ even }\label{especatations}
\\\\
x^{r} (a )_{r}  \frac{{}_1S_0(a +r ;-;x)}{{}_1C_0(a ;-;x)}&\quad\mbox{ if }\;r \;\mbox{ odd }\end{array}\right.\label{espeatations}
\eea
and
\bea
  \langle(\ad)^r \,\ah^r\rangle_o= \frac{x^{r}}{{}_1S_0(a ;-;x)}\left(\frac{d}{dx}\right)^r{}_1S_0(a  ;-;x)=
\left\{\begin{array}{ll}x^{r} (a )_{r} \frac{{}_1S_0(a +r ;-;x)}{{}_1S_0(a  ;-;x)}&\quad\mbox{ if }\;r \;\mbox{ even }\\\\
 x^{r} (a )_{r}  \frac{{}_1C_0(a +r ;-;x)}{{}_1S_0(a ;-;x)}&\quad\mbox{ if }\;r \;\mbox{ odd},\end{array}\right.\label{especatation1s}
\eea
where $x=|z|^2.$\\
For $r=1,$ we deduce   the expectation value of the number operator as
\be 
 \langle N\rangle_e\equiv\langle \ad  \,\ah\rangle_e=ax  \frac{{}_1S_0(a+1  ;- ;x)}{{}_1C_0(a ;-;x)}
\ee
and
\be 
 \langle N\rangle_o\equiv\langle \ad  \,\ah\rangle_o=ax  \frac{{}_1C_0(a+1  ;- ;x)}{{}_1S_0(a;-;x)},\quad x=|z|^2.
\ee
By using 
the expectation value of the operator  $N^2=(  \ad )^2\ah^2+ N$ provided by 
\be
\langle N^2 \rangle_e = a(a+1)x^2\, \frac{  {}_1 C_0 (a+2  ; - ; x)}{ {}_1 C_0 (a  ;- ; x)}
+ ax  \, \frac{  {}_1 S_0 (a  +1  ; -; x)}{ {}_1 C_0 (a  ; - ; x)}
\ee
and
\be
\langle N^2 \rangle_o = a(a+1)x^2\, \frac{  {}_1 S_0 (a +2 ;-; x)}{ {}_1 S_0 (a  ;-; x)}
+ ax\,   \frac{  {}_1 C_0 (a  +1 ; - ; x)}{ {}_1 S_0 (a  ;- ; x)}
\ee
 one readily finds 
\be
 Q_{e}(x)  = x\left((a+1) \frac{  {}_1 C_0 (a+2  ; - ; x)}{ {}_1S_0 (a+1  ;- ; x)}
- a    \frac{  {}_1 S_0 (a  +1  ; -; x)}{ {}_1 C_0 (a  ; - ; x)}\right),
\ee
and
\be
 Q_{o}(x)  = x\left((a+1) \frac{  {}_1S_0 (a+2  ; - ; x)}{ {}_1C_0 (a+1  ;- ; x)}
- a    \frac{  {}_1C_0 (a  +1  ; -; x)}{ {}_1S_0 (a  ; - ; x)}\right),\quad x=|z|^2.
\ee
For  $x<<1$, the Mandel parameter  $ Q_{e}$ is reduced to
\be
 Q_{e}(x)  = 1+o(x^2).
\ee
Therefore, the EGHS   correspond  to the 
  super-Poissonian states. Contrary to that, 
   
\be
 Q_{o}(x)  = -1+o(x^2) <0
\ee
indicating that  the OGHS correspond to the sub-Poissonian states.
The probability of finding $n$ quanta in the   EGHS and OGHS are given by
\be 
 {}_1\mathcal{P}_0^{even}(x,n):=| \langle 2n|a,-,z\rangle_e|^2= \frac{\Big(\frac{a}{2}\Big)_n\Big(\frac{a+1}{2}\Big)_n }{
n!\;\Big(\frac{1}{2}\Big)_n}\frac{x^{2n}}{
{}_2F_1 \left(\begin{array}{c}\frac{a}{2},\frac{a+1}{2}\\
\frac{1}{2}\end{array}\Bigg|\,x^2\right)}, 
\ee
and
\be 
{}_1\mathcal{P}_0^{odd}(x,n):=| \langle 2n+1|a,-,z\rangle_o|^2= \frac{\Big(\frac{a+1}{2}\Big)_n\Big(\frac{a+2}{2}\Big)_n }{
n!\;\Big(\frac{3}{2}\Big)_n}\frac{x^{2n}}{
{}_2F_1 \left(\begin{array}{c}\frac{a+1}{2},\frac{a+2}{2}\\
\frac{3}{2}\end{array}\Bigg|\,x^2\right)},\; x=|z|^2,
\ee
respectively. 
\subsection{The statistical distribution}
The thermodynamics properties are shown to be 
determined by the partition function $Z$ defined by
\be
Z= Tr (e^{-\beta H} )  = \sum_{n=0}^{\infty} \langle n| e^{-\beta H} |n\rangle,
\ee
where $ \beta = 1/kT $.
We assume  the Hamiltonian  to be defined as follows:
\be
\label{gnon}
H:= \omega \ad\ah.
\ee
Now we can compute the partition function for the  generalized oscillator as follows:
\be
Z= \sum_{n=0}^{\infty} \langle n| e^{-\beta H } |n\rangle = \frac {1 }{1- e^{-\beta w }}.
\ee
The statistical distribution of the operator $\hat{O}$ is defined through the formula
\be
\langle \hat{O} \rangle := \frac {1}{Z} Tr ( e^{- \beta H } \hat{O} ).
\ee
Using (\ref{ah:ad}) and (\ref{ah:ad1}), the  mean value of the product of operators $(\ad )^r(\ah )^r$
is  given by:
\be
\label{sawh}
\langle (\ad )^r(\ah )^r \rangle = \frac{(1- e^{-\beta w } )}{(-\omega)^r}\left(\frac{d}{d\beta}\right)^r (1- e^{-\beta w } )^{-1}.
\ee
For $r=1,$ we recover, as expected, the well  known Green function $\langle \ad \ah \rangle $ 
which is the mean occupation number,
 i.e.,
\be
\langle \ad \ah  \rangle = \frac {1}{e^{ w /kT } -1 }.
\ee

\subsection{ Geometry of the states $|a,-,z\rangle_e $ and $|a,-,z\rangle_o $}
 The geometry of a quantum state space can be described by the corresponding metric tensor. This real and positive definite metric is defined on the underlying manifold that the quantum states form, or belong to, by calculating the distance function (line element) between
two quantum states. It is also known as a Fubini-Study metric of the ray space. The knowledge of the quantum metric enables  to calculate quantum mechanical transition probability and uncertainties
\cite{Dezi,HAB}.
The map  $z\longmapsto|a,-,z\rangle_i,\;i=e, o,$ defines a map from the space $C$ of complex numbers onto
a continuous subset of unit vectors in Hilbert space and generates in the latter a two-dimensional surface with the following Fubini-Study metric:
\be 
\label{sama:metric}
 d\sigma^2:= ||d|a,-,z \rangle_i ||^2-| {}_i\langle z,-,a|d|1,-,z\rangle_i |^2.
\ee
 {\bf Proposition 2. } \it 
The Fubini-Study metric  (\ref{sama:metric}) is reduced to
\be 
 d\sigma^2= {}_1W_0^i(x)d\bar z dz,\quad x=|z|^2,
\ee
where 
\be 
 {}_1W_0^{even}(x)=a\frac{d}{dx}\left[x  \frac{{}_1S_0(a+1  ;- ;x)}{{}_1C_0(a ;-;x)}\right]= \frac{d}{dx}\langle N\rangle_e
\ee
and
\be 
 {}_1W_0^{odd}(x)=a\frac{d}{dx}\left[x  \frac{{}_1C_0(a+1  ;- ;x)}{{}_1S_0(a;-;x)}\right]= \frac{d}{dx}\langle N\rangle_o.
\ee
\rm
{\bf Proof.}
Computing $d|a,-,z\rangle_i,\;i=e,o,$ by taking into account the fact that any change of the form
$d|a,-,z\rangle_i=\alpha|a,-,z\rangle_i,\, \alpha\in C$, has zero distance, we get
\be 
 d|a,-,z\rangle_e=\frac{1}{\sqrt{{}_1C_0(a;-;|z|^2)}}\sum_{n=0}^\infty\frac{2nz^{2n-1} }{\sqrt{{}_1\rho_0(2n)}}|2n\rangle \;dz.
\ee
Then,
\bea
 ||d|a,-,z\rangle_e||^2&=&\frac{1}{ {}_1C_0 (a;-;|z|^2)}\sum_{n=0}^\infty\frac{(2n)^2|z|^{2(2n-1)}}{{}_1\rho_0(2n)}\,d\bar z dz
 \cr&=&\frac{1}{{}_1C_0 (a;-;|z|^2)}\left(\sum_{n=0}^\infty\frac{2n|z|^{2(2n-1)}}{{}_1\rho_0(2n)}
+|z|^2\sum_{n=0}^\infty\frac{2n(2n-1)|z|^{2(2n-2)}}{{}_1\rho_0(2n)}\right)\,d\bar z dz
\cr\cr&=&\frac{a\,{}_1S_0 (a+1;-;x)+a(a+1)x\,{}_1C_0 (a+2;-;x)}{{}_1C_0 (a;-;x)}\, d\bar z dz
\cr\cr&=&a\frac{[x\,{}_1S_0 (a+1;-;x)]'}{{}_1C_0 (a;-;x)}\, d\bar z dz
\eea
and
\bea
 |{}_e\langle z,-,a|d|a,-,z\rangle_e|^2&=&
\left|\frac{1}{ {}_1C_0(a;-;x)}\sum_{n=0}^\infty\frac{2n|z|^{2(2n-1)} }{ {}_1\rho_0(2n)} 
 \bar z dz\right|^2
\cr\cr&=&a^2x\left[\frac{ {}_1S_0 (a+1;-;x)}{ {}_1C_0 (a;-;x)}\right]^2 d\bar z dz.
\eea
Therefore,
\bea
 d\sigma^2&=&\left(a\frac{(x\,{}_1S_0 (a+1;-;x))'}{{}_1C_0 (a;-;x)}-
a^2x\,\frac{ {}_1S_0^2(a+1;-;x)}{ {}_1C_0^2(a;-;x)}\right)d\bar{z}dz
\cr&=&\frac{d}{dx}\left[ax\frac{{}_1S_0 (a+1;-;x)}{\,{}_1C_0 (a ;-;x)}\right]d\bar{z}dz\cr
&=& \left(\frac{d}{dx}\langle N\rangle_e\right)d\bar{z}dz,
\eea
where $(\cdot)'$ denotes the derivative with respect to $x$.\\
In the same way, for the OGHS 
\be 
 d\sigma^2= \left(\frac{d}{dx}\langle N\rangle_o\right)d\bar{z}dz,
\ee 
which achieves the proof.

For $x<<1,$ we have
\be
{}_1W_0^{even}(x)=2a(a+1)x+o(x^2),\qquad  {}_1W_0^{odd}(x)=a(a+1)x+o(x^2).
\ee

\section*{Conclusion}

In this work, we  have  investigated a large class of    generalized hypergeometric states
$|p,q,z\rangle$, depending on a complex variable $z$ and two sets of parameters,
$(a_1,\cdots,a_p)$ and $(b_1,\cdots,b_q)$.  Besides, we have defined and analyzed
    even and  odd generalized hypergeometric 
states $|p,q,z\rangle_e$ and $|p,q,z\rangle_o.$ 
We have solved the  moment problem using
the Mellin transform techniques.  For particular values 
of $p$ and $q$,  we have discussed the photon-counting statistics, quantum 
optical properties and geometry of  these states.

\section*{Acknowledgements}
MNH and SA acknowledge   the Abdus Salam International
Centre for Theoretical Physics (ICTP, Trieste, Italy) for its support through the
Office of External Activities (OEA) - \mbox{Prj-15}. The ICMPA
is also in partnership with
the Daniel Iagolnitzer Foundation (DIF), France. One of us (W. S. Chung) was supported
 by the Gyeongsang National University Fund for Professors on Sabbatical Leave, 2006.

\def\JMP #1 #2 #3 {J. Math. Phys. {\bf#1},\ #2 (#3).}
\def\JP #1 #2 #3 {J. Phys. A {\bf#1},\ #2 (#3).}
\def\JPD #1 #2 #3 {J. Phys. D {\bf#1},\ #2 (#3).}
\def\PRL #1 #2 #3 { Phys. Rev. Lett. {\bf#1},\ #2 (#3).}
\def\PR #1 #2 #3 { Phys. Rev. {\bf#1},\ #2 (#3).}
\def\PLA #1 #2 #3 { Phys. Lett. A {\bf#1},\ #2 (#3).}
\def\PLB #1 #2 #3 { Phys. Lett. B {\bf#1},\ #2 (#3).}
\def\PRD #1 #2 #3 { Phys. Rev. D {\bf#1},\ #2 (#3).}
\def\PRA #1 #2 #3 { Phys. Rev. A {\bf#1},\ #2 (#3).}


\section*{References}
\begin{thebibliography}{}
  \bibitem{Klauder&Skagerstam}J. R. Klauder and B. Skagerstam, { \it Coherent States, Applications in Physics and Mathematical Physics}, (World Scientific, Singapore, 1985).

 \bibitem{A.Perelomov} A. Perelomov, { \it Generalized Coherent States and Their Applications}, (Springer, Berlin, 1986).

  \bibitem{E.Schrodinger} E. Schr\"odinger, Naturwissenschaften {\bf 14}, 664 (1926).

 \bibitem{R.Glauber} R. Glauber, \PRA   131   2766  1963

  \bibitem{W.Zhang&D.Feng&R.Gilmore}W. Zhang, D. Feng and R. Gilmore, Rev. Mod. Phys. {\bf 62}, 867 (1990).

  \bibitem{S.Ali&J.Antoine&J.Gazeau} S. T. Ali, J.-P. Antoine and J.-P. Gazeau, {\it Coherent States, Wavelets and Their Generalizations}, (Springer, New York, 2000, $2^{nd}$ edition 2013).

\bibitem{Klauder}J. R. Klauder,   J. Phys. A: Math. Gen. {\bf  29},  L293  (1996).

\bibitem{Littlejohn} R. G. Littlejohn, Phys. Rep. {\bf 138},  193 (1986).

\bibitem{Klauder&PensonS}J. R. Klauder, K. A. Penson and J.-M. Sixdeniers, \PRA 64 013817 2001

  \bibitem{V.Dodonov&I.Malkin&V.Man'ko} V. Dodonov, I. Malkin and V. Man'ko, Physica {\bf 72}, 579 (1974).

   \bibitem{M.Nieto&D.Traux}M. Nieto and D. Traux, \PRL 71 2843 1993

\bibitem{V.Buzek&A.Viiella-Barranco&P.Knight}V. Buzek, A. Viiella-Barranco and P. Knight, \PRA  45  6750 1992

  \bibitem{M.Hillery}M. Hillery, \PRA  36 3796 1987

  \bibitem{J.Hollenhorst}J. Hollenhorst, \PRD   1 3217 1979

\bibitem{Mancini}S.  Mancini, \PLA 233 291 1997

\bibitem{Sivakumar} S. Sivakumar, Phys. Lett. A {\bf  250},  257-262 (1998);  J. Phys. A: Math. Gen. {\bf 33}, 2289-2297 (2000).

\bibitem{Appl}T. Appl and D. H. Schiller,  J. Phys. A: Math. Gen. {\bf 37}, 2731 (2004).

\bibitem{ASK} R. Koekoek and R. Swarttouw,
 {\it The Askey-scheme of hypergeometric orthogonal polynomials and its $q-$analogue},
(Delft University of Technology, Report no. 98-17,  1998).

\bibitem{Marichev} O. I. Marichev, Handbook of Integral Transforms of Higher 
Transcendental Functions, Theory and Algorithmic Tables, (Ellis Harwood, Chichester, 1983).

\bibitem{Akhiezer} N. I. Akhiezer, {\it The Classical Moment Problem and Some Related Questions
in Analysis}, (Oliver and Boyd, London, 1965).

\bibitem{Tamarkin}  J. D. Tamarkin and J. A. Shohat, {\it The Problem of Moments}, (APS, New
York, 1943).
\bibitem{Simon} B. Simon, Adv. Math. {\bf 137}, 82 (1998).

\bibitem{KlauderJR}J. R. Klauder,  K. A. Penson and J.-M. Sixdeniers, Phys. Rev. A {\bf 64} 013817 (2001).

\bibitem{QuesneC} C. Quesne,  Ann. Phys., NY {\bf 293} 147 (2001).

\bibitem{MNSodoga} M. N. Hounkonnou and K. Sodoga, { J. Phys. A: Math. Gen.} {\bf  38}, 7851 (2005).


\bibitem{grad} I. S.  Gradshteyn   and I. M. Ryzhik,   {\it Tables of Integrales, Series and Products,} (Academic, New York, 1980).

\bibitem{OberhettingerF} F. Oberhettinger,  {\it Tables of Mellin Transforms}, (Berlin: Springer, 1974).


\bibitem{DeyFring} S. Dey     and A. Fring,  
\PLA  88 022116 2013

\bibitem{mandel1}L. Mandel,    Opt. Lett. {\bf 4}, 205 (1979).
%
\bibitem{zhangetal} X.-Z. Zhang, Z.-H. Wang, H. Li, Q. Wu, B.-Q. Tang, F. Gao     
  and J.-J. Xu,   Chin. Phys. Lett.   {\bf 25}, 3976  (2008).
%
\bibitem{jpjpg} J.-P. Antoine, J.-P.  Gazeau, P. Monceau, J. R. Klauder 
and K. A. Penson, \JMP 42 2349 2001


\bibitem{MNElvis} M. N. Hounkonnou and E. B. Ngompe 
Nkouankam, J. Phys. A: Math. Theor. {\bf 42} 065202 (2009).

\bibitem{Dezi} M. N.  Hounkonnou      and J. D. Bukweli Kyemba,   
\JMP 51 063518 2010

\bibitem{HAB} M. N. Hounkonnou, S. Arjika and E. Balo\"itcha, arXiv: 1309.6181 [math-ph].

\bibitem{Matos} R. L. de Matos Filho and W. Vogel, \PRA  54 4560 1996

\bibitem{Marmo} V. Man'ko, G. Marmo, E. Sudarshan
and F. Zaccaria, Physica Scripta {\bf 55}, 528 (1997).

\bibitem{Manko&Marmo& Zaccaria}  V. Man'ko, G. Marmo and F. Zaccaria, quant-ph/9703020 (1997).

\bibitem{Manko}  V. Man'ko, \PLA 228 29 1997

\bibitem{Junker&Roy}  G. Junker and P. Roy, \PLA   257 113 1997

\bibitem{SMancini} S. Mancini, \PLA   233 291 1997

\bibitem{BRoy}  B. Roy, \PLA  249 25 1998

\bibitem{Roy&Roy}  B. Roy and P. Roy, J. Opt. B { \bf  1}, 341 (1999).

\bibitem{Roy&Roy2}  B. Roy and P. Roy, \PLA 257 264  1999

\bibitem{Sivakumar1} S. Sivakumar, \PLA  250 257 1998

\end{thebibliography}
\end{document}